\documentstyle[aps,epsf,rotate,citesort,multicol]{revtex}

\begin{document}

\draft
\title{Non-L\'evy Distribution of Commodity Price Fluctuations}

\author{Kaushik Matia$^{1}$, Luis A. Nunes Amaral$^1$, Stephen P.
Goodwin$^2$ and H. Eugene~Stanley$^1$} 

\address{
$^1$Center for Polymer Studies and Department of Physics,
  Boston University, Boston, MA 02215\\
$^2$ BP Upstream, Chertsey Road, Sunbury-on-Thames, Middlesex TW16 7LN\\
}

\maketitle

\begin{abstract}

Price fluctuations of commodities like cotton and wheat are thought to
display probability distributions of returns that follow a L\'evy stable
distribution. Recent analysis of stocks and foreign exchange markets
show that the probability distributions are not L\'evy stable, a
plausible result since commodity markets have quite different features
than stock markets. We analyze daily returns of 29 commodities over
typically 20 years and find that the distributions of returns decay as
power laws with exponents $\alpha$ which have values $\alpha > 2$,
outside the L\'evy-stable domain. We also find that the amplitudes of
the returns display long-range time correlations, like stocks, while the
returns themselves are uncorrelated for time lags $\approx$ 2 days, much
larger than for stocks ( $\approx$ 4 min).

\end{abstract}


\begin{multicols}{2}


The study of economic markets has recently become an area of active
research for physicists, for various reasons. First, markets
constitute complex systems for which the variables characterizing the
state of the system---i.e., the price of the goods, the number of
trades, the number of agents are easily quantified. Second, because of
the importance of markets there is a large amount of data that can be
accessed.

Much of the research interest of physicists has concentrated on stocks
~\cite{Stocks}, stock averages~\cite{Index}, foreign exchange rates
~\cite{Foreign-exchange} and other markets~\cite{Other-markets}. A number
of key empirical findings have been established: (i) The distribution of
logarithmic price changes is approximately symmetric and decays with
power law tails with an exponent $\alpha + 1 \approx 4$ for the
probability density function~\cite{Stocks,Index,Foreign-exchange}. (ii)
The price changes are uncorrelated beyond rather short time scales
\cite{note2}; and (iii) the amplitude of the price changes have
long-range correlations, specifically, the correlations decay as a power
law with an exponent $\gamma \approx 1/3$~\cite{Stocks}.

Perhaps the most intriguing aspect of these empirical findings is that
they appear to be universal, that is individual US stocks appear to
conform to these ``laws''~\cite{Stocks}, as do German stocks
~\cite{Lux}, and Australian stocks~\cite{Allison}.  Moreover, market
indices such as the S\&P 500, the Dow Jones, the NIKKEI,
the Hang Seng or the Milan index~\cite{Index} also
obey these same laws.  Furthermore, similar results are found for the
most traded currency exchange rates such as the US dollar vs.\ the
Deutsch mark, or the US dollar vs.\ the Japanese yen
~\cite{Dacorogna}. The ``universal'' nature of the statistics of daily
returns is remarkable since the markets described above are quite
different in their details. The observed universality thus suggests a
rather striking similarity in the underlying mechanisms.

Unlike stock and foreign exchange markets, commodity markets have
received less attention
~\cite{Commodity1,Commodity2,Commodity3,Commodity4,Commodity5,Commodity6}. Contrary
to heavily traded stocks or currencies---which have a somewhat abstract
character because they (i) have an almost ``elastic'' response to
changes in demand, (ii) do not require storage, and (iii) are not
``consumed''--- commodities are physical products that are traded
because they (i) cannot be produced at will, (ii) require physical
storage, and (iii) are needed for some purpose.  For example, one needs
gasoline to run a car, heating oil to heat a home, or electricity to
light an office.

Since stock markets and commodity markets differ in so many respects,
one might imagine that commodity prices show larger fluctuations than
stock prices. In fact, probability distributions of the returns of
commodities such as cotton and wheat have been reported
~\cite{Commodity1,Commodity2,Commodity3} to be L\'evy stable that have
power law decaying tails with exponents in the range $0 <
\alpha < 2$, which implies the probability of a large gain or loss is
larger than if the distribution were a power law with $\alpha > 2$
~\cite{Stocks,Index}. Here we address the question whether the scaling
of commodities is statistically distinguishable from that of stocks. We
study the statistical properties of the price fluctuations for 29
commodities~\cite{platt} and compare the results with the statistical
properties of daily returns in stock markets.
 
\begin{figure}
\narrowtext
\vspace*{0.6cm}
\centerline{
\epsfysize=0.75\columnwidth{\rotate[r]{{\epsfbox{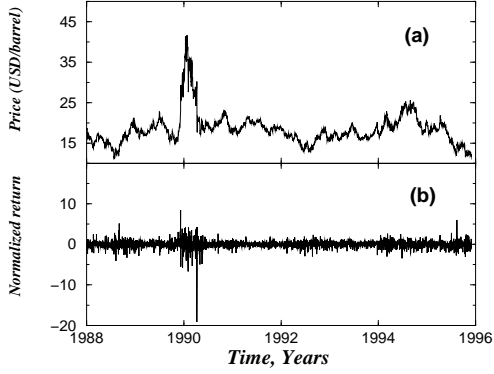}}}}
}
\vspace*{0.6cm}
\caption{ (a) Prices $S_i(t)$ for a typical commodity, high-sulphur fuel
oil HSFO arg [Table~\protect\ref{table}]. (b) Normalized returns
$g_i(t)$ defined by Eq.~(\protect\ref{e.return}). Note the large
fluctuations of up to $20\sigma$, which would have probability $\approx
e^{-200}$ for a Gaussian distribution.}
\label{f-return}
\end{figure}

\begin{figure}
\narrowtext
\vspace*{0.6cm}
\centerline{
\epsfysize=0.45\columnwidth{\rotate[r]{{\epsfbox{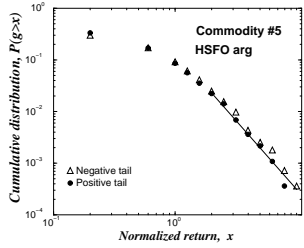}}}}
}
\vspace*{0.6cm}
\caption{Cumulative distributions of positive and negative returns for
high-sulphur fuel oil (HSFO arg) [Table~\protect\ref{table}].  The tails
of the distributions shown have power law decays with an exponent
$\alpha_5 = 2.6 \pm 0.3$ for the negative tail and $\alpha_5 = 2.9 \pm
0.06$ for the positive tail, where the exponents and the error bars are
estimated by Hill's method ~\protect\cite{Hill75}.}
\label{f-cum}
\end{figure}

We define the normalized price fluctuation (``return'')
\begin{equation}
g_i(t) \equiv \frac{\ln S_i(t+\Delta t)- \ln S_i(t)}{\sigma_i}  ,
\label{e.return}
\end{equation}
where $\Delta t = 1$ day, $i = 1,2,..,29$ indexes the commodities,
$S_i(t)$ is the price and $\sigma_i $ is the standard
deviation. Figure~\ref{f-return} displays the price and corresponding
returns of a typical commodity, high-sulphur fuel oil [Table
~\ref{table}]. Figure~\ref{f-cum} shows that the probability
distributions $P(g_i > x)$ for both positive and negative tails decay as
power laws with exponents that we cannot distinguish statistically,

\begin{equation}
P(g_i > x) \sim \frac{1}{x^{\alpha_i}}
\label{e.cum_prob}
\end{equation} 
where $\alpha_i$ is outside the L{\'e}vy stable domain
$0<\alpha_i<2$. 

\begin{figure}
\narrowtext
\vspace*{0.6cm}
\centerline{
\epsfysize=0.55\columnwidth{\rotate[r]{{\epsfbox{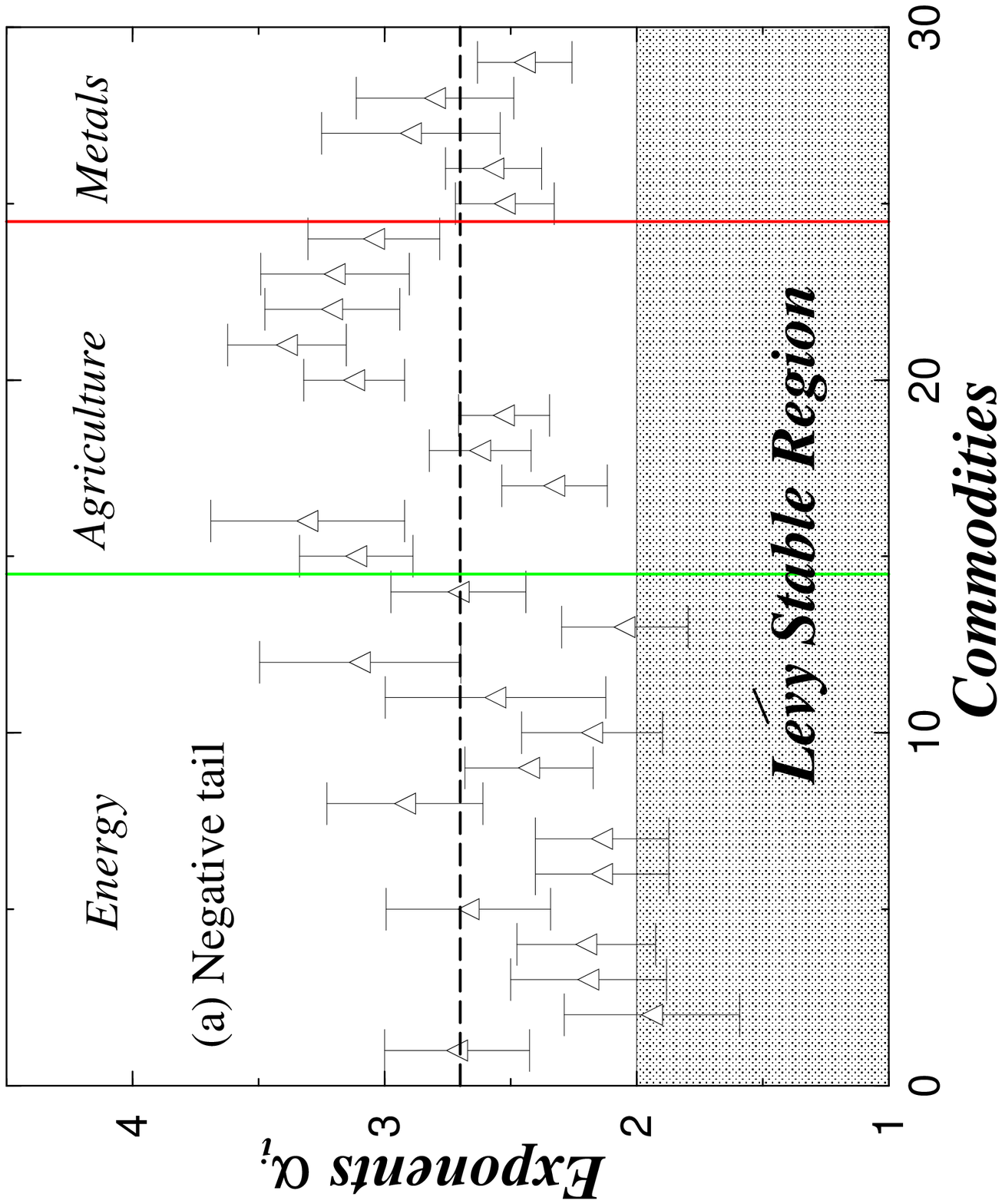}}}}
\epsfysize=0.55\columnwidth{\rotate[r]{{\epsfbox{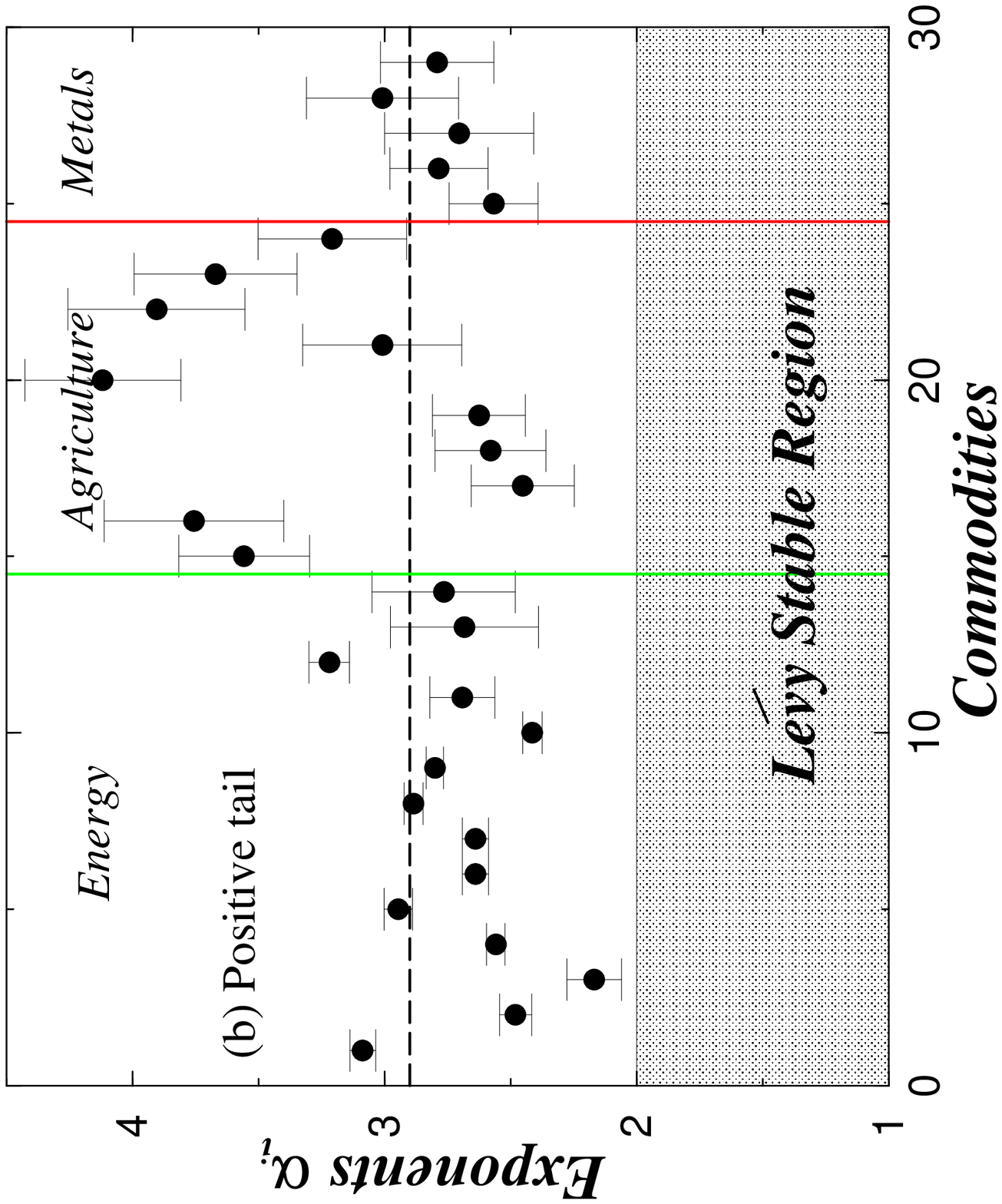}}}}
}
\vspace*{0.6cm}
\caption{ (a) Exponents $\alpha_i$ of the negative tail and (b) the
positive tail, where $i = 1,2,..,29$ indexes the 29 commodities analyzed
[Table~\protect\ref{table}]. We employ Hill's
method~\protect\cite{Hill75} to estimate the exponent $\alpha_i$ of each
probability distribution in the range $ x \ge x_{\rm cutoff}$ where we
choose $x_{\rm cutoff} = 2$. The dashed lines show the average values
$\alpha = (\sum_{i=1}^{29} \alpha_i)/29$ . Shaded regions indicate the
range of L\'evy stable exponents, $\alpha<2$.}
\label{f-exponents} 
\end{figure}

Figure ~\ref{f-exponents} displays Hill estimates ~\cite{Hill75} of
$\alpha_i$ for all 29 commodities, calculated for $x$ above a cutoff
value $x_{\rm cutoff}$, which is estimated by evaluating $\alpha_i$ starting
from different trial values of $x_{\rm cutoff}$, and choosing $x_{\rm cutoff}$
as the minimum $x$ value above which $\alpha_i$ does not change
significantly. Based on our analysis we choose for all 29 commodities
the same value $x_{\rm cutoff} = 2 $. Note that for daily returns the number
of data points beyond $x_{\rm cutoff} = 2 $ is typically 50-200. The average
exponent is $\alpha = (\sum_{i=1}^{29} \alpha_i)/29 =2.9 \pm 0.08$ for
the positive tail and $\alpha=2.7 \pm 0.07$ for the negative
tail~\cite{note}.

\begin{figure}
\narrowtext
\vspace*{0.6cm}
\centerline{
\epsfysize=0.55\columnwidth{\rotate[r]{{\epsfbox{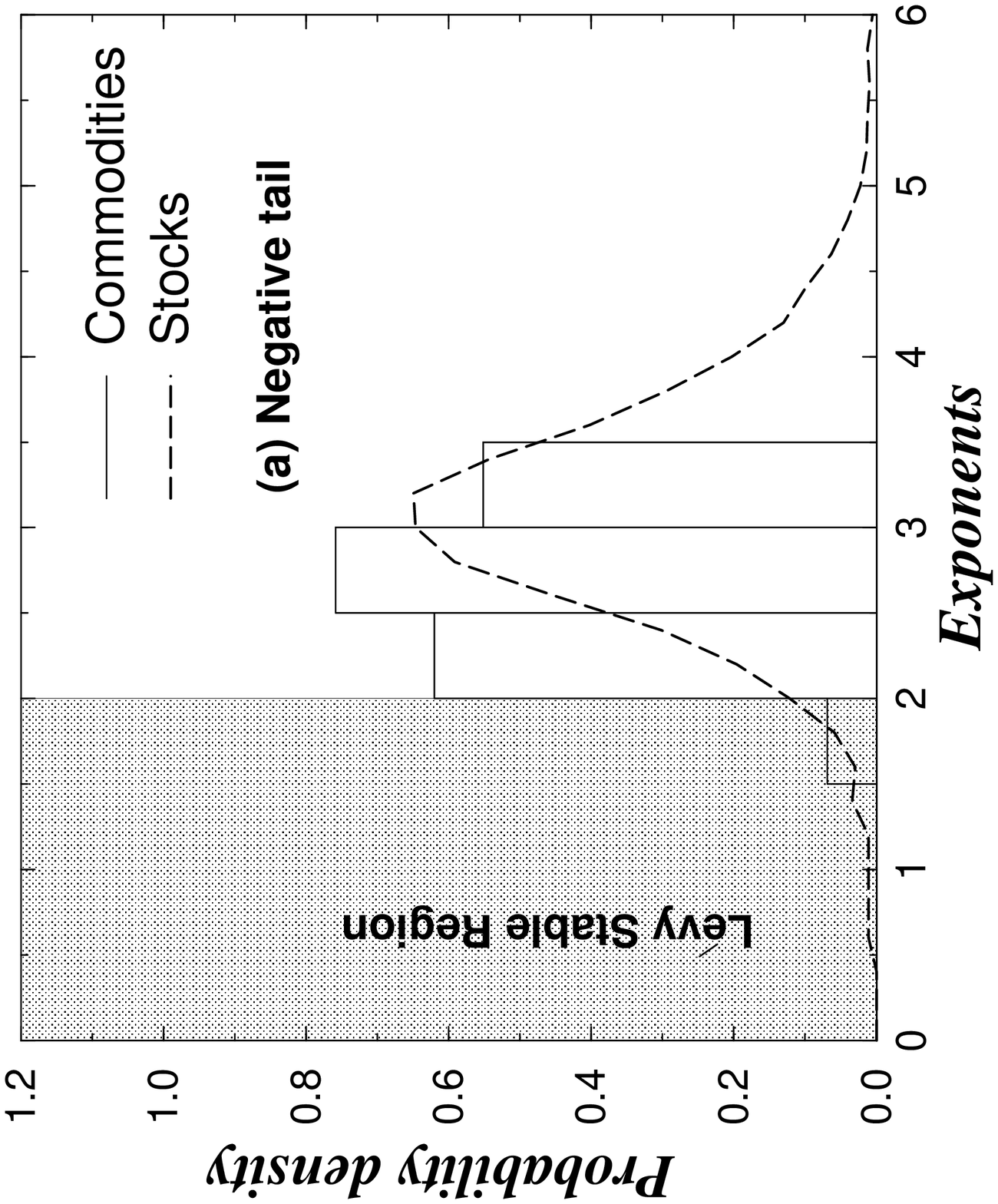}}}}
\epsfysize=0.55\columnwidth{\rotate[r]{{\epsfbox{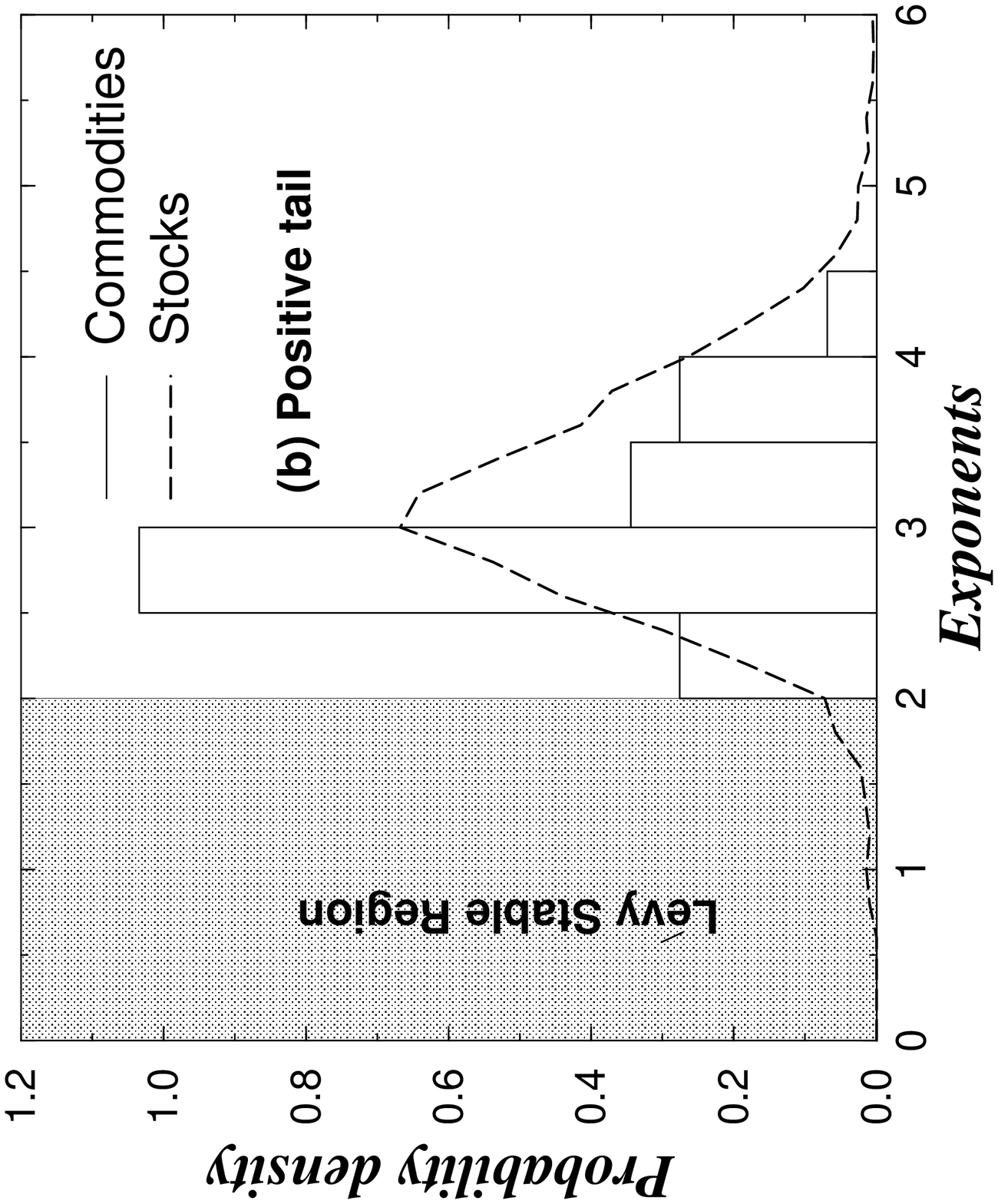}}}}
}
\vspace*{0.6cm}
\caption{ Probability density function of (a) the negative tail
exponents and (b) the positive tail exponents evaluated for the 7128
stocks and the 29 commodities analyzed. Observe that for both stocks and
commodities the mean exponent is around 3.0, outside the L\'evy stable
region $0< \alpha < 2$.}
\label{f-histo-exponents} 
\end{figure}

Next we compare our calculations of $\alpha_i$ with exponents $\alpha_i$
of daily returns evaluated for 7128 stocks from the CRSP
database~\cite{Stocks,crsp}. We choose stocks in the same time period as
the 29 commodities analyzed, and compute tail exponents of $P(x)$ by the
same procedure. Figures ~\ref{f-histo-exponents}(a) and
~\ref{f-histo-exponents}(b) compare the probability density functions of
tail exponents for both commodities and stocks. The pdf for stocks is
same as is reported in ~\cite{Stocks}, and the exponents are outside the
L\'evy stable region for both stocks and commodities.

\begin{figure}
\narrowtext
\vspace*{0.6cm}
\centerline{
\epsfysize=0.45\columnwidth{\rotate[r]{{\epsfbox{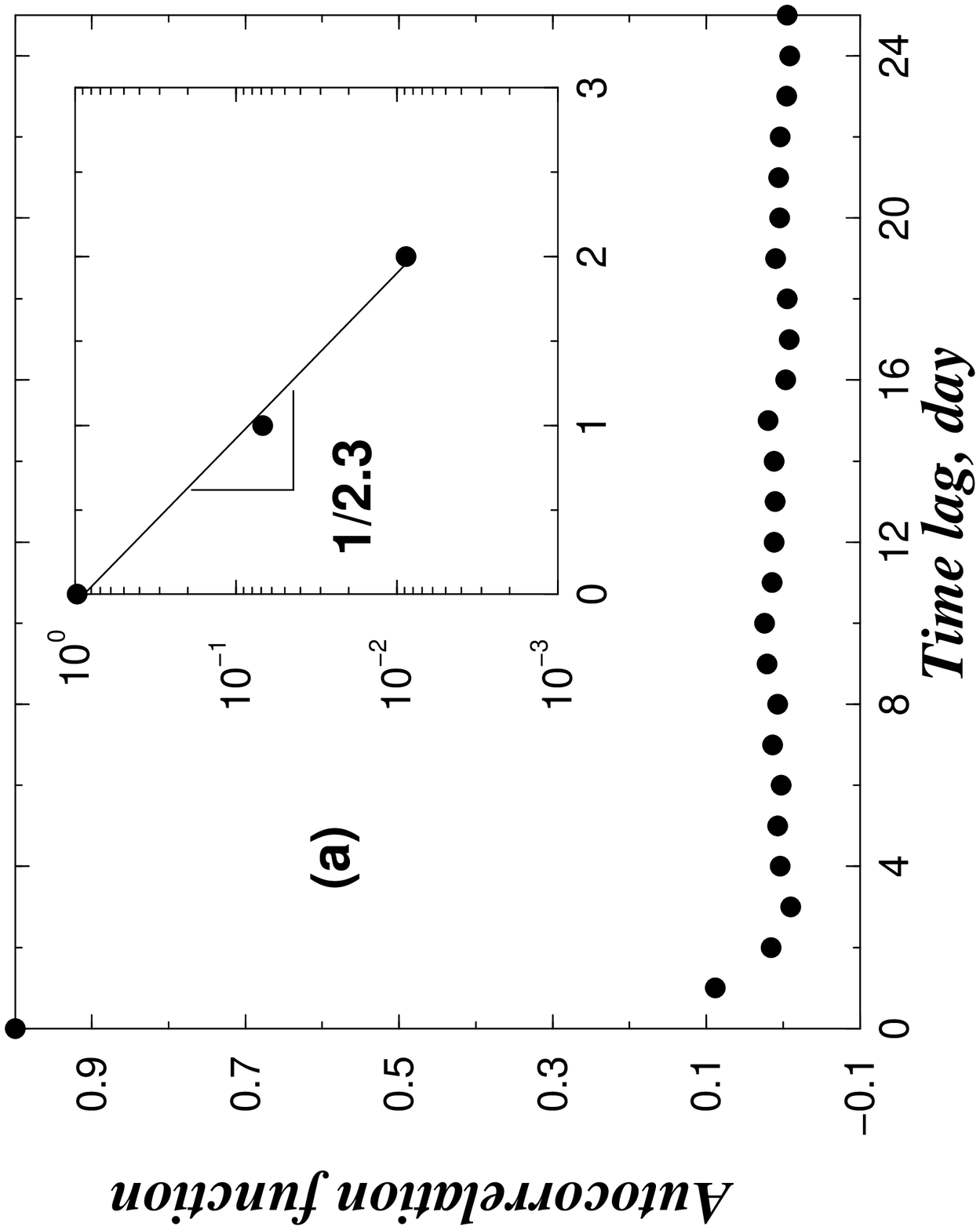}}}}
\epsfysize=0.47\columnwidth{\rotate[r]{{\epsfbox{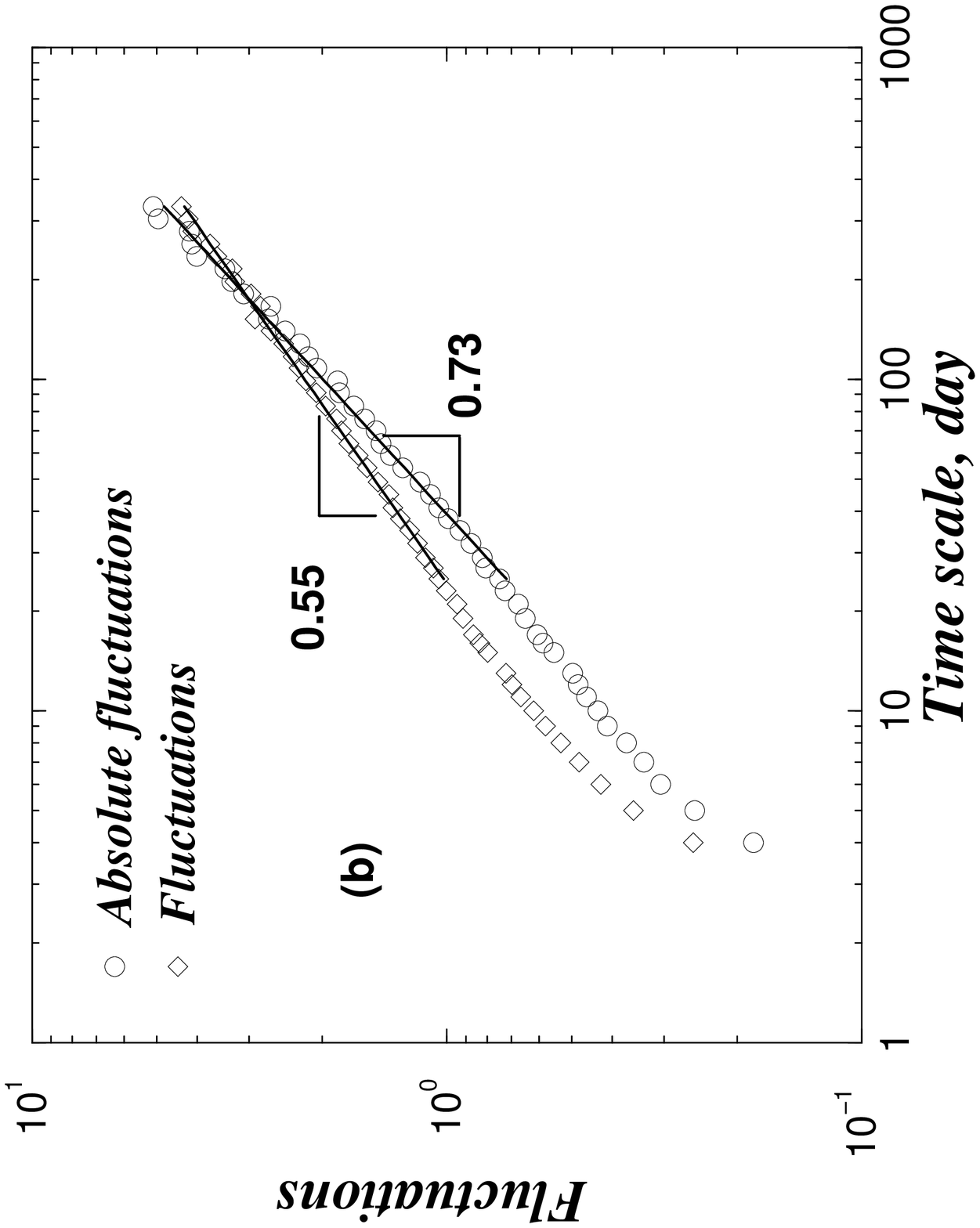}}}}
}
\vspace*{0.6cm}
\caption{ (a) Autocorrelation function averaged over all 29
commodities. The autocorrelation function displays an exponential decay
with a decay constant of 2.3 days (see inset). (b) The fluctuations
$F(n)$ of returns $g_i$ and absolute value of returns $|g_i|$ averaged
over all 29 commodities. The plot for $g_i$ is consistent with the
exponential decay of Fig. 5(a). The exponent of $\hat\alpha = 0.73 \pm
0.05$ for $|g_i|$ implies long-range power law correlation, a feature
also seen for stock markets where $\hat\alpha \approx
0.8$~\protect\cite{Stocks}.}
\label{f-dfa} 
\end{figure}

We next discuss time correlations of returns. Figure~\ref{f-dfa}(a)
displays the autocorrelation function $<g_i(t)g_i(t+\tau)>$ averaged
over the 29 commodities. We observe that this averaged autocorrelation
function $C(\tau)$ ceases to be statistically different from zero for
time lags $\tau$ of 3~days or more. To further quantify time
correlations, we use the detrended fluctuation analysis (DFA)
method~\cite{Peng}. The DFA method calculates fluctuations $F(n)$ in a
time window of size $n$, and then plots $F(n)$ versus $n$. The slope
$\hat\alpha$ in a log-log plot gives information about the correlations
present. If $C(\tau) \sim \tau^{-\gamma}$ then $\hat\alpha = (2
-\gamma)/2 $, while if $C(\tau) \sim e^{-\tau/\tau_c}$ then $\hat\alpha
= 1/2$~\cite{Peng}. We find in Fig.~\ref{f-dfa}(b) that $\hat\alpha =
0.55 \pm 0.05$, consistent with the exponential decay of
Fig.~\ref{f-dfa}(a). We also observe that $|g_i|$, the absolute value of
returns (one measure of volatility), are power law correlated with
$\hat\alpha = 0.73 \pm 0.05$, which implies a power law decay of the
autocorrelation of the absolute value of returns with $\gamma = 0.54 \pm
0.1$.



We thank S. V. Buldyrev, X. Gabaix, P. Gopikrishnan, V. Plerou,
A. Schweiger and especially P. King for helpful discussions and
suggestions and BP/Amoco for financial support.




\newpage
\renewcommand{\baselinestretch}{1}
\begin{table}
\caption{Description of commodities analyzed}
\begin{tabular}{ccp{9.0cm}cp{1.5cm}}
Index   & Symbol  & Description                                         & Period   & \# points \\
\hline
\hline
1    & Brent      & Crude oil (UK Standard cf WTI in the US)            & 1/88--8/98 & 2770 \\  
\hline
2   & BUTANE	  & Butane chemical used as fuel                        & 2/93--8/98 & 1433 \\
\hline
3   & Gasoil      & Gas oil                                             & 1/88--7/93 & 1433 \\
\hline
4      & HFO      & Heavy fuel oil                                      & 1/88--8/98 & 2770 \\
\hline
5  & HSFO arg     & High-sulfur fuel oil from Arabian Gulf              & 1/88--8/98 & 2770 \\
\hline
6  & HSFO New	  & High-sulfur fuel oil transported by New York barge  & 1/88--8/98 & 2770 \\
\hline
7 & Kero New	  & Kerosene transported by New York barge              & 1/88--8/98 & 2770 \\
\hline
8 & LSFO New     & Low-sulfur fuel oil transported by New York barge   & 1/88--8/98 & 2770  \\ 
\hline
9 & LSFO NYH	  & Low-sulfur fuel oil traded at New York Harbour      & 1/88--8/98 & 2770 \\
\hline
10  & Nap Med      & Naphtha from Mediterranean (used for feed-stock)     & 1/88--8/98 & 2770  \\
\hline
11  & Nap New	  & Naphtha transported by New York barge               & 1/88--4/95 & 1897  \\
\hline
12 & Prem unl	  & Premium unleaded automobile gasoline                & 6/92--8/98 & 1619 \\
\hline
13       & CL     & Crude Oil           & 1/83--8/99  & 4164 \\ 
\hline
14       & HO     & Heating Oil         & 9/79--8/99  &  4999 \\
\hline
15        & C     & Corn                & 6/69--8/99  & 7593 \\
\hline
16       & CT     & Cotton              & 6/69--8/99  &  7593 \\  
\hline
17       & FC     & Feeder Cattle       & 7/79--8/99  &  5051 \\ 
\hline
18       & KC     & Coffee              & 6/69--8/99  &  7593 \\ 
\hline
19       & O     & Oats                & 6/69--8/99  &  7593 \\ 
\hline
20        & S     & Soybeans            & 6/69--8/99  &  7593 \\ 
\hline
21       & SB     & Sugar               & 1/80--8/99  &  4909 \\ 
\hline
22       & W      & Wheat               & 1/75--8/99  &  6186 \\
\hline
23       & LH     & Live Hogs           & 6/69--8/99  &  7593 \\
\hline
24       & PB     & Pork Bellies        & 6/69--8/99  &  7593 \\ 
\hline 
25       & GC     & Gold                & 6/69--8/99  &  7593 \\ 
\hline
26       & HG     & Copper High Grade   & 1/71--8/99  &  7195 \\ 
\hline
27       & PA     & Palladium           & 6/87--8/99  &  3039 \\ 
\hline
28       & PL     & Platinum            & 6/87--8/99  &  3039 \\ 
\hline
29       & SI     & Silver              & 6/69--8/99  &  7593 \\
\end{tabular}
\vspace{0.2cm}
\label{table}
\end{table}
\normalsize




\end{multicols}

\begin{references}

\bibitem{Stocks} P.~Gopikrishnan {\it et al.} Eur. Phys. J. B {\bf 3},
139 (1998); Y.~Liu {\it et al.} Phys. Rev. E {\bf 60}, 1390
(1999); V.~Plerou {\it et al.} Phys. Rev. E {\bf 60}, 6519 (1999).

\bibitem{Index}  R.~N.~Mantegna and H.~E. Stanley, Nature {\bf 376}, 46
(1995); P.~Gopikrishnan {\it et al.} Phys. Rev. E {\bf 60}, 5305
(1999).

\bibitem{Foreign-exchange} M.~M. Dacorogna, U.~A. M\"uller,
R.~J. Nagler, R.~B. Olsen, and O.~V.  Pictet, J. Int'l Money and Finance
{\bf 12}, 413 (1993).

\bibitem{Other-markets} G.~Weisbuch, A.~Kirman, D.~Herreiner, Econ J.
{\bf463}, 411 (2000); J.~P.~Nadal, G.~Weisbuch, O.~Chenevez and
A.~Kirman, in {\it Advances in Self-Organization and Evolutionary
Economics}, edited by J. Lesourne and A. Orlian (Economica, London,
1998), p. 149

\bibitem{note2} For stock markets it is seen that the autocorrelation
function decays exponentially to the noise level within a time scale of
4 minutes~\cite{Stocks,Index}.

\bibitem{Lux} 
T.~Lux, Applied Financial Economics {\bf 6}, 463 (1996).

\bibitem{Allison} A.~Allison and D.~Abbott, {\it Unsolved Problems of
Noise}, edited by D.~Abbott and L.~Kish (AIP proceedings, Melville, New
York, 2000).

\bibitem{Dacorogna}
 U~ A. M\"uller, M.~M. Dacorogna, R.~B. Olsen, O.~V. Pictet, M.
~Schwarz, and C.~Morgenegg, J. Banking and Finance {\bf 14}, 1189 (1995).

\bibitem{Commodity1} B. B. Mandelbrot, J. Business {\bf 36}, 394 (1963).

\bibitem{Commodity2} W. Working, Wheat Studies {\bf 9}, 187 (1933).

\bibitem{Commodity3} W. Working, Wheat Studies of the Stanford Food
Institute {\bf 2}, 75 (1934).

\bibitem{Commodity4} R. Weron, Physica A {\bf 285}, 127 (2000). 

\bibitem{Commodity5} B. M. Roehner, Eur. Phys. J. B {\bf 8}, 151 (1999). 

\bibitem{Commodity6} B. M. Roehner, Eur. Phys. J. B {\bf 13}, 175 (2000).

\bibitem{platt} See {\tt http://www.platts.com} and {\tt http://finance.yahoo.com}

\bibitem{Hill75} B.~M.~Hill, Ann. Stat. {\bf3}, 1163 (1975).

\bibitem{note} If we assume that all 29 commodities follow the same
distribution then we can aggregate the data, in this way we find $\alpha
= 2.9 \pm 0.6$ for the positive and $\alpha = 2.7 \pm 0.05$ for the negative tail.

\bibitem{crsp} See {\tt http://www.crsp.com}

\bibitem{Peng} C.~K.~Peng, S.~V.~Buldyrev, S.~Havlin, M.~Simons,
H.~E.~Stanley, A.~L.~Goldberger, Phys. Rev. E {\bf 49}, 1685 (1994).

\end{references}
\end{document}